\documentclass[aps,prl,twocolumn,showpacs]{revtex4}
\usepackage{graphicx}
\usepackage{dcolumn}
\usepackage{amsmath}
\begin{document}

\title{Proper Scaling of the Anomalous Hall Effect}
\author{Yuan Tian}
\author{Li Ye}
\author{Xiaofeng Jin}
\email[Corresponding author, email: ]{xfjin@fudan.edu.cn}
\affiliation{Surface Physics Laboratory and Physics Department,
Fudan University, Shanghai 200433, China}
\date{\today}

\begin{abstract}
Working with epitaxial films of Fe, we succeeded in independent
control of different scattering processes in the anomalous Hall
effect. The result appropriately accounted for the role of phonons,
thereby clearly exposing the fundamental flaws of the standard plot
of the anomalous Hall resistivity versus longitudinal resistivity. A
new scaling has been thus established that allows an unambiguous
identification of the intrinsic Berry curvature mechanism as well as
the extrinsic skew scattering and side-jump mechanisms of the
anomalous Hall effect.
\end{abstract}


\maketitle

Shortly after the discovery of the Hall effect, in 1880 Edwin Hall
further observed in ferromagnetic metals an additional large
contribution besides the ordinary one, which is now called the
anomalous Hall effect (AHE) - one of the most prominent phenomena
existing in magnetic materials [1]. While the ordinary Hall effect
has been well understood as a result of the Lorentz force deflecting
the charge carriers, the mechanism of the AHE has remained
controversial despite the long history of research, because its rich
phenomenology defies the standard classification methodology,
prompting conflicting reports claiming the dominance of various
processes [2-12]. Recently it again attracts great attention because
of its natural connection to the spin Hall effect and quantum spin
Hall effect [13, 14].

In ferromagnets, the transverse resistivity has two contributions:
one is ordinary and is proportional to the applied magnetic field;
the other is anomalous and is normally proportional to the
magnetization [8, 9]. It is often written as
\begin{equation}
\rho_{xy}=r_{0}H+r_{a}M\equiv\rho_{h}+\rho_{ah}
\end{equation}
where $r_{0}$ and $r_{a}$ are coefficients that characterize the
strength of the ordinary and anomalous Hall resistivity $\rho_{h}$
and $\rho_{ah}$ , respectively. It has long been believed that
$\rho_{ah}$ should be a function of the longitudinal resistivity
$\rho_{xx}$ with well-defined material material dependent
parameters, i.e., $\rho_{ah}=f(\rho_{xx})$ .

However, despite the tremendous amount of experiments, a proper
scaling between $\rho_{ah}$ and $\rho_{xx}$ has not yet been
established. In general, $\rho_{ah}$ could exhibit four types of
behavior: (a) $b\rho_{xx}^{2}$ (e.g. Fe) [15-18], (b)
$a\rho_{xx}+b\rho_{xx}^2$ (e.g. Co) [19, 20], (c)
$b\rho_{xx}^{\alpha}$ (e.g. Ni) with $1<\alpha<2$ [21], and (d)
$a\rho_{xx}$ (e.g. ultra-pure Ni at low temperature) [22]. A unified
picture that can explain all the diversifying experimental facts is
currently absent.

Theoretically, Karplus and Luttinger first proposed that the
spin-orbit interaction together with the interband mixing resulted
in an intrinsic anomalous velocity in the direction transverse to
the electric field [2], which gave $\rho_{int}\propto\rho_{xx}^{2}$.
This intrinsic contribution to the AHE has been recently confirmed
in the language of Berry phase [10-12]. However, Smit suggested that
the skew scattering at impurities was responsible for the AHE, which
gave $\rho_{sk}\propto\rho_{xx}$ [3]. Berger further proposed that
another impurity-induced mechanism, the side-jump, could also give
the $\rho_{sj}\propto\rho_{xx}^{2}$ relation [5]. In contrast,
recent first principles electronic band structure calculations based
on the Berry phase interpretation suggested (although implicitly)
that it is the Karplus-Luttinger intrinsic contribution rather than
any impurity-induced extrinsic ones that plays the dominant role in
the AHE [19, 23-25]. Apparently these predictions are contradictory
to each other, while embarrassingly a direct experimental
identification is still lacking. In a recent review by Sinitsyn [9],
it is shown that the total anomalous Hall conductivity
($\sigma_{ah}$) consists of five different microscopic
contributions; they can be further divided into the above three
categories in terms of the experimental identification by transport
measurement, i.e.,
$\sigma_{ah}=\sigma_{int}+\sigma_{sk}+\sigma_{sj}$, where
$\sigma_{int}$ is the intrinsic Karplus-Luttinger contribution and
is irrelevant of any impurity, $\sigma_{sk}$ is the extrinsic skew
scattering that depends on the impurity density, and $\sigma_{sj}$
the generalized extrinsic side-jump (including not only the
conventional side-jump but also the contributions from intrinsic
skew scattering and the anomalous distribution) that does not depend
on the impurity density. It is a great challenge to separate the
roles of intrinsic and extrinsic contributions in experiment.

Rising to these challenges we have developed a new experimental
strategy that goes beyond the existing paradigms. Usually the
longitudinal resistivity is taken as a quantity characteristic only
of the material, so traditionally besides temperature, $\rho_{xx}$
was varied only by changing the impurity concentration of the
material. However, such an approach has the unavoidable defect that
it would in fact modify not only the extrinsic but also the
intrinsic contributions in the AHE, often complicating the
interpretation of the experimental results. Instead, we are tuning
the resistivity $\rho_{xx}$ by varying the film thickness of
ultrathin layer of Fe, an idea similar to the control of the
coefficient of viscosity for a moving fluid in a thin tube by
varying its diameter; the only difference is that the former deals
with electrons while the latter with molecules. It is this novel
approach that helps appropriately account for the role of phonons in
the impurity-originated scatterings, resulting in the proper scaling
of the AHE:
$\sigma_{ah}=-(a_{sk}\sigma_{xx0}^{-1}+b_{sj}\sigma_{xx0}^{-2})\sigma_{xx}^{2}+\sigma_{int}$
or
$\rho_{ah}=(a_{sk}\rho_{xx0}+b_{sj}\rho_{xx0}^{2})-\sigma_{int}\rho_{xx}^{2}$
. Here $\rho_{xx0}$ is the residual resistivity caused by defects in
crystal, $a_{sk}$ and $b_{sj}$ are material dependent constants for
the skew scattering and side-jump, respectively. This new scaling
enables us to separate out the intrinsic Karplus-Luttinger
contribution from the various impurity-originated extrinsic
contributions, and further singles out its dominant role for the AHE
as $\sigma_{xx}^{2}\rightarrow 0$ . We finally develop a unified
physical picture that can explain all the previous experimental
results.

\begin{figure}
\center
\includegraphics[width=8.5cm]{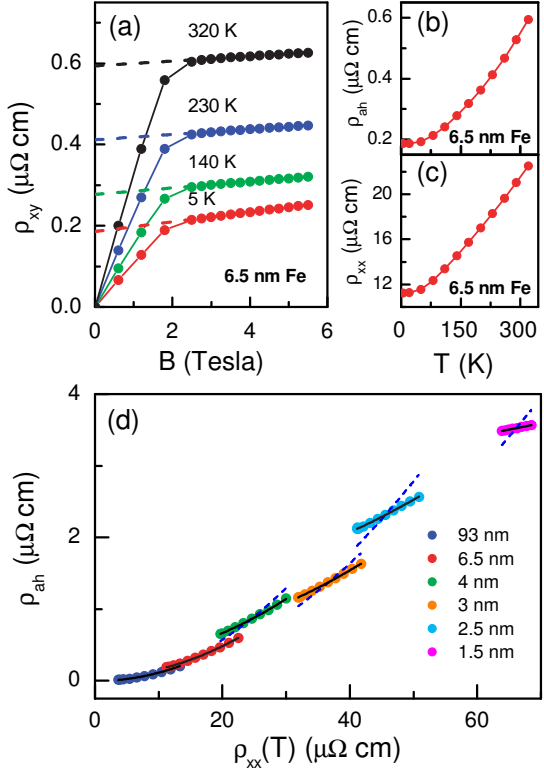}
\caption{(color online) (a) Experimentally measured representative
$\rho_{xy}$ vs $H$ curves for 6.5 nm film, from which $\rho_{ah}$
can be obtained. (b) and (c) $\rho_{ah}$ and $\rho_{xx}$ as
functions of temperature for 6.5 nm film, respectively. (d)
$\rho_{ah}$ and $\rho_{xx}$ for various film thicknesses. The dashed
blue and solid black lines are fitting results with
$\rho_{ah}=b'\rho_{xx}^{2}$  and
$\rho_{ah}=a'\rho_{xx0}+a''\rho_{xxT}+b\rho_{xx}^{2}$,
respectively.}
\end{figure}

Fe films were grown on undoped GaAs(001) at 300 K by molecular beam
epitaxy (MBE), and were capped with 4 nm thick MgO to prevent
oxidation in air. The detailed experimental setup was described
elsewhere [26, 27]. The films were patterned into the form of a
standard Hall bar along [110] with the magnetic field along [001].
The transport measurements were carried out in a physical property
measurement system (Quantum Design PPMS-9T system). The
magnetoresistance in Fe films at 5 Tesla is smaller than 0.5\%, and
the magnetization in Fe films thicker than 1 nm has its bulk value
and changes little within the temperature range of 5 K-320 K.

The experimental relation between the anomalous Hall resistivity
$\rho_{ah}$ and the longitudinal resistivity $\rho_{xx}$ is
established through the following procedure. Fig. 1(a) shows several
representative sets of $\rho_{xy}$ vs $H$ curves measured at
different temperatures between 5 K and 320 K for a 6.5 nm thick Fe
film. $\rho_{ah}(T)$ is then obtained as the zero field
extrapolation of the high field data as shown in the figure, and is
displayed in Fig. 1(b). $\rho_{xx}(T)$ was measured simultaneously
and is shown in Fig. 1(c). From Fig. 1(b) and 1(c), $\rho_{ah}$
versus $\rho_{xx}$ curve for the 6.5 nm Fe film can be deduced and
is displayed in Fig. 1(d), together with data for other thicknesses
varying between 1.5 nm and 93 nm.

This experimentally established $\rho_{ah}=f(\rho_{xx})$ at
different film thicknesses provides an opportunity to unveil the
phonon contribution to the AHE - a long-standing controversial issue
[3, 6, 28]. Before going to the detailed data analysis, we recall
first some very basics about the anomalous Hall effect. According to
Ohm's law, there exists a general relation between the anomalous
Hall resistivity and the anomalous Hall conductivity:
$\rho_{ah}=-\sigma_{ah}/(\sigma_{xx}^{2}+\sigma_{ah}^{2})$ or
$\sigma_{ah}=-\rho_{ah}/(\rho_{xx}^{2}+\rho_{ah}^{2})$, which is
material independent and is valid for each independent mechanism of
the AHE. Since $\sigma_{ah}<<\sigma_{xx}$, $\rho_{ah}<<\rho_{xx}$,
and $\sigma_{xx}=\rho_{xx}^{-1}$, this can be further simplified to:
 $\rho_{ah}=-\sigma_{ah}\rho_{xx}^{2}$ or $\sigma_{ah}=-\rho_{ah}\sigma_{xx}^{2}$.
It immediately follows that different mechanisms are additive both
in the anomalous Hall conductivity and resistivity, i.e.,
$\sigma_{ah}=\sum\sigma_{j}$ and $\rho_{ah}=\sum\rho_{j}$ - an
important fact that is quite often overlooked. Specifically, a
constant $\sigma_{int}$ therefore
$\rho_{int}=-\sigma_{int}\rho_{xx}^{2}$ is expected for any given
material [2, 11, 12]; on the other hand, it was generally adopted
without serious justification that the skew scattering and side-jump
were expressed as $\rho_{sk}=a_{sk}\rho_{xx}$ and
$\rho_{sj}=b_{sj}\rho_{xx}^{2}$ respectively. Following this line of
reasoning, it seems quite natural why the formula
$\rho_{ah}=a_{sk}\rho_{xx}+b\rho_{xx}^{2}$ has been widely used to
investigate the AHE and separate various contributions (e.g., [18,
19, 25]). But we are going to expose the fundamental flaws of this
formula in the following, then establish a new scaling for the AHE.

\begin{figure}
\center
\includegraphics[width=8.5cm]{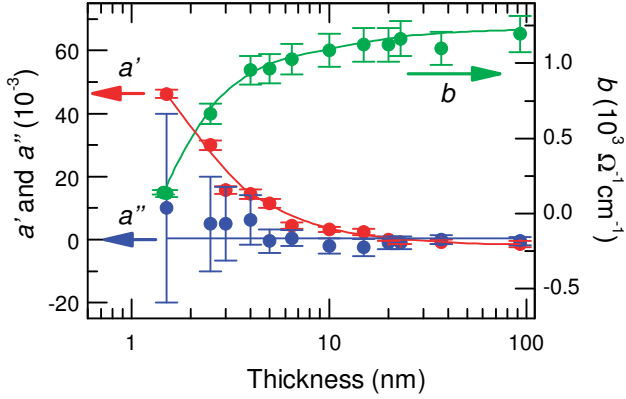}
\caption{(color online). Red, blue, and green dots represent
respectively the parameters $a'$, $a''$ and $b$. Solid curves are
guide to the eye.}
\end{figure}

Following the observation in bulk Fe reported in Ref. [15, 16], we
used $b'\rho_{xx}^{2}$ to fit the data in Fig. 1(d), but found the
significant deviation when the Fe films were thinner than 6.5 nm, as
clearly seen by the dashed lines in the figure. This means that the
skew scattering must be considered in ultrathin Fe films although it
is negligible in the bulk material. We therefore use the routinely
accepted formula $\rho_{ah}=a_{sk}\rho_{xx}+b\rho_{xx}^{2}$ to fit
the data, but found $a_{sk}$ is not a constant, showing strong
thickness dependence (not shown here). However since the impurity
density concentration and the boundary condition remain the same for
different film thickness, it is known that $\rho_{xx}$ decreases as
the film thickness increases, but it is puzzling why $a_{sk}$ should
be thickness dependent or $\rho_{xx}$ dependent ($a_{sk}(\rho_{xx})$
). Unlike the experiments with variable impurity where
$a_{sk}(\rho_{xx})$ might happen, here in this new approach the fact
that $a_{sk}$ is not a constant strongly implies the improper
scaling of the formula. We start to question why the skew scattering
should be included as $\rho_{sk}=a_{sk}\rho_{xx}$. According to the
Matthiessen's rule $\rho_{xx}=\rho_{xx0}+\rho_{xxT}$ as seen in Fig.
1(c), $\rho_{sk}=a_{sk}\rho_{xx}$ implies
$\rho_{sk}=a_{sk}\rho_{xx0}+a_{sk}\rho_{xxT}$, i.e., the phonons
would contribute (via $\rho_{xxT}$) equally to the skew scattering
just like the defects in crystal (via $\rho_{xx0}$). This is
certainly a very strong assumption and must be verified. In order to
achieve this we first assume they contribute unequally to the skew
scattering, i.e., $\rho_{sk}=a'\rho_{xx0}+a''\rho_{xxT}$ , then try
to determine whether $a'$ and $a''$ are identical by fitting the
data in Fig. 1(d) with
$\rho_{ah}=a'\rho_{xx0}+a''\rho_{xxT}+b\rho_{xx}^{2}$ using the
experimentally measured $\rho_{ah}$, $\rho_{xx0}$, $\rho_{xxT}$, and
$\rho_{xx}$ for different film thicknesses, similar to those
obtained for the 6.5 nm film in Fig. 1(b) and 1(c). The fits are
shown by the solid lines in Fig. 1(d), and the fitting parameters of
$a'$, $a''$ and $b$ are presented in Fig. 2. It is evident from Fig.
2 that $a'$ and $a''$ are very different and actually $a''\approx
0$. However, it seems that the same puzzle - $a'$ is thickness
dependent - still exists; it turns out that the two situations are
fundamentally different, which becomes clear in the following.

Similarly, the same concern exists for the side-jump as well, i.e.,
why it should be included as $\rho_{sj}=b_{sj}\rho_{xx}^{2}$ in the
widely used formula, unlike in the intrinsic Karplus-Luttinger case
where $\sigma_{int}$ is a real constant so that
$\rho_{int}=-\sigma_{int}\rho_{xx}^{2}$ follows directly from the
general relation between the anomalous Hall conductivity and
resistivity. If the side-jump
$\rho_{sj}=b_{sj}\rho_{xx0}^{2}+b_{sjT}\rho_{xxT}^{2}$ (it does not
affect the following results even if there exists cross terms of
$\rho_{xx0}\rho_{xxT}$) is considered in parallel and together with
the skew scattering $\rho_{sk}=a_{sk}\rho_{xx0}+a_{skT}\rho_{xxT}$,
then without losing the generality what we should verify is
$\rho_{ah}=(a_{sk}\rho_{xx0}+a_{skT}\rho_{xxT})+(b_{sj}\rho_{xx0}^{2}+b_{sjT}\rho_{xxT}^{2})-\sigma_{int}\rho_{xx}^{2}$
. It seems unreasonable from the first glance to work out so many
parameters from Fig. 1(d) alone; however, the situation is
dramatically changed when we regroup the formula in the form of
$\rho_{ah}=(a_{sk}+b_{sj}\rho_{xx0})\rho_{xx0}+(a_{skT}+b_{sjT}\rho_{xxT})\rho_{xxT}-\sigma_{int}\rho_{xx}^{2}$,
which has exactly one-to-one correspondence with the formula
$\rho_{ah}=a'\rho_{xx0}+a''\rho_{xxT}+b\rho_{xx}^{2}$ used to check
the skew scattering above. Apparently we can reinterpreted the same
Fig. 2 as the following: $a''\approx 0$ is a direct experimental
justification that phonons actually contribute little to the overall
extrinsic AHE as compared to that of the defects in crystal, either
the contributions from the skew scattering and side-jump cancels
each other, or they are both negligible; $b$ is nothing but the
negative anomalous Hall conductivity $b=-\sigma_{int}$, which is
fully developed and becomes almost saturated
($b=-\sigma_{int}\approx1.1\times10^{3} \Omega^{-1}cm^{-1}$) as the
film thickness reaches 4 nm and above; the thickness dependent $a'$
 in Fig. 2 simply indicates that the extrinsic AHE contains
contributions not only from the skew scattering ($a_{sk}\rho_{xx0}$)
but also from the side-jump ($b_{sj}\rho_{xx0}^{2}$) so that a
thickness dependence $a'$ should be expected as
$a'=(a_{sk}+b_{sj}\rho_{xx0})$. Following this line of logic, a new
scaling of $\rho_{ah}=f(\rho_{xx},\rho_{xx0})$ rather than the
traditional $\rho_{ah}=f(\rho_{xx})$ is proposed for the AHE:
$$\rho_{ah}=(a_{sk}\rho_{xx0}+b_{sj}\rho_{xx0}^{2})-\sigma_{int}\rho_{xx}^{2}
\ \ \ \ \eqno{(2A)}$$
$$\sigma_{ah}=-(a_{sk}\sigma_{xx0}^{-1}+b_{sj}\sigma_{xx0}^{-2})\sigma_{xx}^{2}+\sigma_{int}
\eqno{(2B)}$$ Here, $\sigma_{int}$, $a_{sk}$ and $b_{sj}$ are all
material dependent constants.

\begin{figure}
\center
\includegraphics[width=8.5cm]{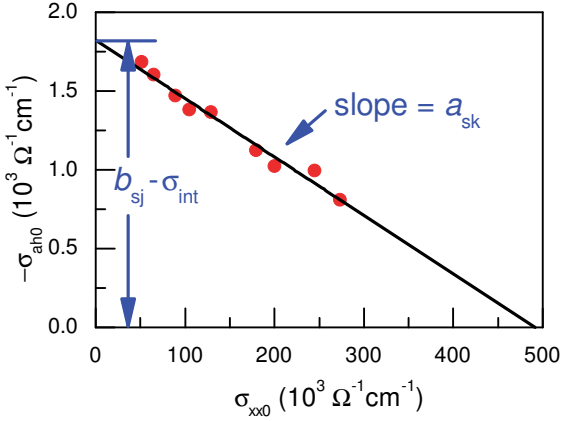}
\caption{(color online). The red dots represent the $-\sigma_{ah0}$
vs $\sigma_{xx0}$ relation using the experimentally measured raw
data at 5 K for different film thicknesses. The black curve is the
fitting by
$-\sigma_{ah0}=a_{sk}\sigma_{xx0}+(b_{sj}-\sigma_{int})$.}
\end{figure}

To confirm indeed $a'=(a_{sk}+b_{sj}\rho_{xx0})$, i.e., the
extrinsic anomalous Hall resistivity
$\rho_{ext}=a_{sk}\rho_{xx0}+b_{sj}\rho_{xx0}^{2}$ or anomalous Hall
conductivity
$\sigma_{ext}=-(a_{sk}\sigma_{xx0}^{-1}+b_{sj}\sigma_{xx0}^{-2})\sigma_{xx}^{2}$
 as in Eq. 2A and
2B, we plot in Fig. 3 $-\sigma_{ah0}$ vs $\sigma_{xx0}$ as shown by
the red dots, using the experimental raw data measured at 5 K for
different film thicknesses. At this temperature, besides
$\rho_{xx0}=\sigma_{xx0}^{-1}$ we also have
$\sigma_{ah}\approx\sigma_{ah0}$ and
$\sigma_{xx}\approx\sigma_{xx0}$, so that Eq. 2B becomes:
$-\sigma_{ah0}=a_{sk}\sigma_{xx0}+(b_{sj}-\sigma_{int})$, by which
the experimental data can be well fitted as seen from the black line
in the figure, meanwhile the corresponding constants are extracted
as: $a_{sk}=-3.7\times10^{-3}$ and
$(b_{sj}-\sigma_{int})=1.8\times10^{3} \Omega^{-1}cm^{-1}$.
Recalling the previous result $-\sigma_{int}\approx1.1\times10^{3}
\Omega^{-1}cm^{-1}$ from Fig. 2, we furthermore get the side-jump
constant $b_{sj}\approx0.7\times10^{3} \Omega^{-1}cm^{-1}$. It is
evident that at low temperatures the side-jump $b_{sj}$ is not
negligible comparing with the Karplus-Luttinger intrinsic term
$-\sigma_{int}$; on the other hand, the change of
$a_{sk}\sigma_{xx0}$ as a function of $\sigma_{xx0}$ for the samples
explored in this experiment as seen in Fig. 3 can be as large as
$1.0\times10^{3} \Omega^{-1}cm^{-1}$, almost the same magnitude of
the $-\sigma_{int}$ value, thus is also not small at all. In
addition, the negative sign of $a_{sk}=-3.7\times10^{-3}$ indicates
that the skew scattering contributes to the AHE in Fe in the
opposite direction as the side-jump and the Karplus-Luttinger terms
do. Therefore in principle it could exceed them and becomes dominant
at low temperature for samples with larger $\sigma_{xx0}$,
explaining the striking and long puzzled phenomenon in which the
anomalous Hall resistivity of Fe would change sign simply as the
temperature is lowered as observed earlier [16], which was unable to
understand with the simple $\rho_{ah}=b'\rho_{xx}^{2}$ term.

Instead of the data fitting as done in Fig. 1(d) and Fig. 2, we are
going to single out now the Karplus-Luttinger intrinsic contribution
from the extrinsic contributions in a much more straightforward and
transparent way. Fig. 4 shows the $\sigma_{ah}$ versus
$\sigma_{xx}^{2}(T)$ plot using the experimental raw data, each
curve corresponding to a specific film thickness with variable
temperatures between 5 K and 290 K. This figure contains some
important information about AHE, which has been hidden too long.
First of all, the linear relationship between $\sigma_{ah}$ and
$\sigma_{xx}^{2}$ confirms in an elegant way that phonons do
contribute little to the skew scattering and side-jump in the AHE,
as predicted by Eq. 2B, otherwise a linear relationship of
$\sigma_{ah}$ versus $\sigma_{xx}$ would be expected from the widely
used but flawed formula $\rho_{ah}=a_{sk}\rho_{xx}+b\rho_{xx}^{2}$.
Then, as $\sigma_{xx}^{2}$ goes to zero in the figure, the anomalous
Hall conductivity $\sigma_{ah}$ converges to essentially the same
but nonzero value. This converged value is nothing but exactly the
long searched Karplus-Luttinger intrinsic or the Berry curvature
contribution, i.e.,
$\sigma_{ah}=(\sigma_{int}+\sigma_{sk}+\sigma_{sj})\rightarrow\sigma_{int}$
when $\sigma_{xx}\rightarrow0$ according to Eq. 2B. To flesh out
this critical point, we believe that as $\sigma_{xx}^{2}$
approaching zero (i.e. in the high temperature limit for metals),
the extrinsic terms ($\sigma_{sk}+\sigma_{sj}$) attributed to the
impurity scattering ought to be washed out by the random and
incoherent phonon scattering, therefore shrunk to zero as seen in
Fig. 4; in the meantime,  the intrinsic one of $\sigma_{int}$
originated from the electronic structure of the material should be
the only robust one that is usually less sensitive to temperature.
In addition, this converged value of
$-\sigma_{int}\approx(1.1\pm0.1)\times10^{3} \Omega^{-1}cm^{-1}$ at
$\sigma_{xx}^{2}=0$ for films thicker than 4 nm not only agrees very
well to the ($b=-\sigma_{int}$) result of Fig. 2, but also to that
from bulk Fe whisker measured at room temperature [16, 24] as marked
by the red dashed line, which demonstrates unambiguously that it is
the Karplus-Luttinger intrinsic rather than any extrinsic mechanisms
that plays the dominant role for the AHE in bulk Fe at room
temperature and higher. It should also be pointed out that for films
thinner than 4 nm the $-\sigma_{int}$ value decreases as seen in
Fig. 2 (not shown in Fig. 4), presumably due to the finite-size or
quantum-well modification (in ultrathin film of Fe) to the bulk
electronic band structure.

\begin{figure}
\center
\includegraphics[width=8.5cm]{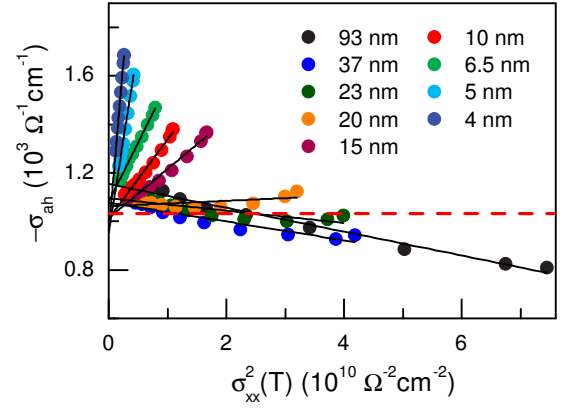}
\caption{(color online). $-\sigma_{ah}$ vs $\sigma_{xx}^{2}$ for Fe
films with various thicknesses. The black lines are linear fitting
of corresponding data. The red dashed line corresponds to the
$-\sigma_{ah}$ value obtained from iron whisker. }
\end{figure}

With the new proper scaling of the anomalous Hall effect, Eq. 2A and
2B, we can now finally unify all the aforementioned diverse
experimental results in literature. First, it is likely that type
(a) as defined in the introduction corresponds to situations in
which either the material dependent parameters $a_{sk}$ and $b_{sj}$
happen to be very small, or the measurements were carried out at
temperatures where $\rho_{xx0}\ll\rho_{xx}$. Second, we have
reanalyzed the data of type (b) using Eq. 2A, and found that the new
scaling can indeed describe better those experimental results.
Third, type (d) actually belongs to a special case of Eq. 2A where
the temperature was fixed very low (so $\rho_{xx}\approx\rho_{xx0}$)
and the sample was ultra-pure (so $\rho_{xx0}^{2}\ll\rho_{xx0}$),
thus the second and third terms in the equation are negligible
compared to the first, leading to an equation of $\rho_{ah}\approx
a_{sk}\rho_{xx0}$. Fourth, type (c) corresponds to a nontrivial case
of Eq. 2A, where the intrinsic conductivity $\sigma_{int}$ itself is
sensitive to temperature in the range of interest [23, 24]. However,
because of the absence of the Matthiessen's rule in semiconductors,
it is not trivial whether the new scaling established here in Eq. 2
applies to the extrinsic AHE in magnetic semiconductors as well,
although nothing is against it yet. Last but not least, if an
experiment is carried out in a constant longitudinal current mode
for an almost perfect ferromagnetic crystal at extremely low
temperature, the anomalous Hall voltage is expected to be
essentially zero according to Eq. 2A, but the anomalous Hall current
is expected to be finite as seen from Eq. 2B.

We believe that the new result presented here opens wide
possibilities to manipulate in a controlled way either intrinsic or
extrinsic or both effects to meet certain application purposes in
future spintronics devices.

This work was supported by MSTC (No. 2006CB921303 and No.
2009CB929203) and NSFC (No. 10834001).

\end{document}